# Electricity generation using molten salt technology

Charles Osarinmwian

School of Chemical Engineering and Analytical Science, University of Manchester, Oxford Road, Manchester M13 9PL

The anodic release of carbon dioxide gas in the molten salt Hall-Héroult process can be used to power a turbine for electricity generation. The application of this new concept in molten salt reprocessing in the nuclear industry is considered because it could facilitate the suitability of carbon dioxide cycles to certain types of nuclear reactor. The theoretical power of 27.8 MW generated by a molten salt Hall-Héroult reactor is comparable with a next-generation biomass plant that sources low-grade wood fuel to generate renewable heat and power using CHP steam turbine technology.

## Power generation

The concept of generating electricity from molten salt processes will lead to the utilization of carbon dioxide cycles in these processes. For nuclear energy to be sustainable, new large-scale fuel cycles will be required that may include fuel reprocessing[1]. In recent years, stable anodes based on highly electronically conducting alkali ruthenates have been developed for molten salt electrolytes for high temperature batteries (with proposed designs based on Hall-Héroult reactors)[2]. In the molten salt Hall-Héroult process, used for industrial scale aluminum production, the reactor typically consists of a steel shell lined with refractory alumina, carbon and a thermal insulator. Electric current enters the reactor through the carbon-based anode and flows through the alumina-containing molten salt, to the molten aluminum deposited at the bottom of the reactor. There are two types of consumable carbon-based anode used: monolithic self-baked (Soderberg) and prebaked, where reactors containing prebaked anodes are more efficient and differ only in fabrication and stub connection[3].

The deposition of alumina on the bottom of the reactor, caused by poor dissolution and diffusion of alumina in the molten salt, creates alumina concentration gradients that generate non-uniform current distributions in the Hall-Héroult reactor. Installing sloped 0.57-0.61 cm high drainable carbon cathodes coated with titanium diboride to offer wettability of aluminum could help inhibit alumina deposition at the bottom of the reactor[3,4]. The dissolution and diffusion of alumina is mainly dependent on circulation patterns in the molten salt that are driven by anodic gas bubbles[5]. These bubbles are evolved from anode reactions and lower the effective conductivity of the molten salt; this influences the total reactor resistance and the local anodic current distribution where polarization effects in cathodic reactions contribute much less to the anodic overvoltage[6].

*Anode reactions*

$$Al_2O_2F_6^{2-} + 2F^- + C \rightarrow \mathbf{CO_2} + 2AlF_4^- + 4e^-$$

$$Al_2O_2F_4^{2-} + 4F^- + C \rightarrow \mathbf{CO_2} + 2AlF_4^- + 4e^-$$

*Cathode reactions*

$$AlF_6^{3-} + 3e^- \rightarrow Al + 6F^-$$

$$AlF_4^- + 3e^- \rightarrow Al + 4F^-$$

On-line monitoring of any changes in electrical conductivity induced by anodic gas bubbles could be performed non-invasively using electrical impedance tomography (Fig. 1). Although imaging the electrical conductivity distribution in the reactor would mark a paradigm shift in molten salt technology, several problems will need to be addressed (e.g. imaging is limited by the difficulty of quantifying the reliability of tomographic images and poor image distinguishability caused by the poor location of surrounding electrodes)[7,8]. It is important to note that the primary applications of electrical impedance tomography are in medicine and physiology[9].

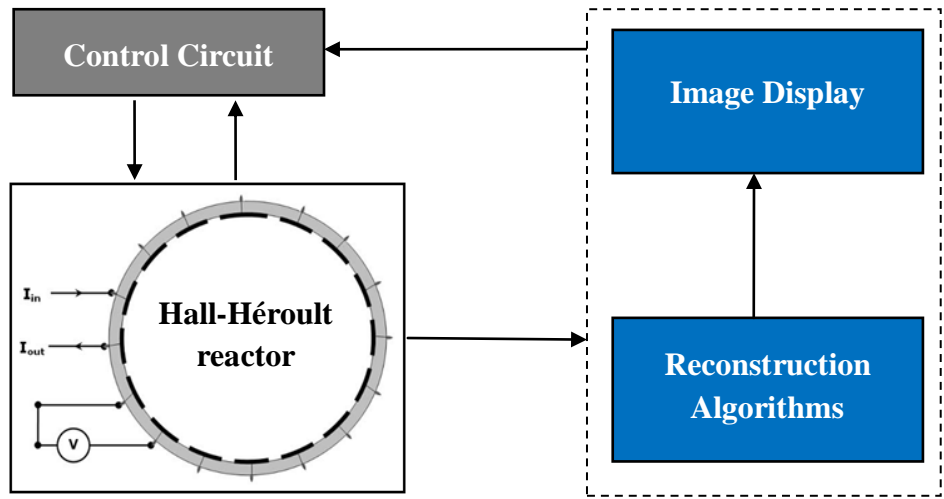

**Figure 1: Schematic of an on-line electrical impedance tomography system for a Hall-Héroult reactor. Electrical impedance data is generated by injecting current into the Hall-Héroult reactor from a current source and then measuring the voltage between electrodes that externally surround the reactor.**

To determine a theoretical power delivered by a turbine and the outlet temperature of the carbon dioxide gas stream from this turbine, an isentropic expansion of the gas stream originating from the anode passes at 527 °C and 80 bar across the turbine at 1 bar. Temperature control of the gas stream is crucial since the supercritical carbon dioxide cycle is well suited to any type of nuclear reactor with a core outlet temperature above ~500 °C[10]. The outlet temperature is determined from $S(T_1, P_1) = S(T_2, P_2)$ which indicates that the total entropy change of the gas stream $\Delta S$ as a function of temperature $T$ and pressure $P$ (inlet: subscript 1; outlet: subscript 2) is zero. The entropy of the system is a sum of the ideal gas contribution $S^{ig}$ with the residual contribution $S^{res}$ where the entropy change of the ideal gas is

$$S^{ig}(T_2, P_2) - S^{ig}(T_1, P_1) = \int_{T_1}^{T_2} \frac{dT}{T} C_p(T) - R \ln \frac{P_2}{P_1}$$

$$= \int_{T_1}^{T_2} \frac{dT}{T} \left( A + BT + \frac{D}{T^2} \right) - R \ln \frac{P_2}{P_1}$$

$$= A \ln \frac{T_1}{T_2} + B(T_2 - T_1) - \frac{D}{2}\left(\frac{1}{T_2^2} - \frac{1}{T_1^2}\right) - R \ln \frac{P_2}{P_1} \qquad 1$$

where $C_p$ is the ideal gas specific heat capacity of carbon dioxide, $A$, $B$, $D$ are ideal gas heat capacity coefficients (Table 1) and $R$ is the universal gas constant. The residual entropy is computed from the Peng-Robinson equation of state:

$$P = \frac{RT\rho}{1 - \rho b} - \frac{a(T)\rho^2}{1 + 2\rho b - \rho^2 b^2} \qquad 2$$

where

$$a(T) = \frac{0.45724 R^2 T_c^2}{P_c}\left\{1 + m\left[1 + \left(\frac{T}{T_c}\right)^{1/2}\right]\right\}^2 \qquad 3$$

$$m = 0.37464 + 1.54226\omega - 0.26992\omega^2 \qquad 4$$

$$b = \frac{0.07780 R T_c}{P_c} \qquad 5$$

and $\omega$ is the accentric factor (Table 1), $\rho$ is the density of carbon dioxide gas, $T_c$ and $P_c$ are the critical temperature and pressure of carbon dioxide gas respectively.

**Table 1: Properties of carbon dioxide**

| Ideal gas heat capacity coefficients | |
|---|---|
| $A = 5.457$; $B = 1.045 \times 10^{-3}$ K$^{-1}$; $D = -1.157 \times 10^5$ K$^2$ | |
| **Critical properties** | **Accentric factor** |
| $T_c = 304.2$ K; $P_c = 73.8$ bar | $\omega = 0.225$ |

In order to compute the residual entropy, a derivative with respect to temperature is taken for an expression of the residual Helmholtz free energy $A^{res}$. The residual Helmholtz free energy is computed by integrating the compressibility factor $Z$ as follows

$$\frac{A^{res}}{NRT} = \int_0^\rho \frac{d\rho}{\rho}(Z-1) = \frac{P}{\rho RT} \qquad (6)$$

where $N$ is the number of moles. Combining Eqs. 2 and 6 gives

$$\begin{aligned}
\frac{A^{res}}{NRT} &= \int_0^\rho \frac{d\rho}{\rho}\left[\frac{\rho b}{1-\rho b} - \frac{a(T)\rho}{RT(1+2\rho b - \rho^2 b^2)}\right] \\
&= \int_0^\rho \frac{d\rho}{\rho}\left[\frac{b}{1-\rho b} - \frac{a(T)}{RT(\sqrt{2}+1-\rho b)(\sqrt{2}-1+\rho b)}\right] \\
&= \int_0^\rho \frac{d\rho}{\rho}\left[\frac{b}{1-\rho b} - \frac{a(T)}{2\sqrt{2}RT(\sqrt{2}+1-\rho b)} - \frac{a(T)}{2\sqrt{2}RT(\sqrt{2}-1+\rho b)}\right] \\
&= \left[-\ln(1-\rho b) + \frac{a(T)}{2\sqrt{2}bRT}\ln(\sqrt{2}+1-\rho b) - \frac{a(T)}{2\sqrt{2}bRT}\ln(\sqrt{2}-1+\rho b)\right]_0^\rho \\
&= -\ln(1-\rho b) + \frac{a(T)}{2\sqrt{2}bRT}\ln\frac{(\sqrt{2}+1-\rho b)}{(\sqrt{2}-1+\rho b)} - \frac{a(T)}{2\sqrt{2}bRT}\ln\frac{(\sqrt{2}+1)}{(\sqrt{2}-1)} \\
&= -\ln(1-\rho b) + \frac{a(T)}{2\sqrt{2}bRT}\ln\left[\frac{1-(\sqrt{2}-1)\rho b}{1+(\sqrt{2}+1)\rho b}\right] \qquad (7)
\end{aligned}$$

The residual entropy is computed from the residual Helmholtz free energy by

$$\begin{aligned}
S^{res}(T,P) &= -\frac{\partial A^{res}}{\partial T} \\
&= -\frac{\partial}{\partial T}\left\{-RT\ln(1-\rho b) + \frac{a(T)}{2\sqrt{2}b}\ln\left[\frac{1-(\sqrt{2}-1)\rho b}{1+(\sqrt{2}+1)\rho b}\right]\right\}
\end{aligned}$$

$$= R\ln(1-\rho b) - \frac{1}{2\sqrt{2}b}\frac{\partial a(T)}{\partial T}\ln\left[\frac{1-(\sqrt{2}-1)\rho b}{1+(\sqrt{2}+1)\rho b}\right] \qquad (8)$$

where

$$\frac{\partial a(T)}{\partial T} = a(T_c)\frac{\partial}{\partial T}\left\{1+m\left[1-\left(\frac{T}{T_c}\right)^{1/2}\right]\right\}^2 \qquad ()$$

$$= -\frac{a(T_c)m}{T_c(T/T_c)^{1/2}}\left\{1+m\left[1-\left(\frac{T}{T_c}\right)^{1/2}\right]\right\} \qquad (9)$$

Hence, the simultaneous equations are of the form:

$$S^{res}(T_2,P_2) = -\left[S^{ig}(T_2,P_2)-S^{ig}(T_1,P_1)\right] + S^{res}(T_1,P_1) \qquad (10)$$

$$P_2 = \frac{RT_2\rho_2}{1-\rho_2 b} - \frac{a(T_2)\rho_2^2}{1+2\rho_2 b - \rho_2^2 b^2} \qquad (11)$$

After solving Eq. 2 for the density of the inlet gas stream, Eqs. 10 and 11 are solved to give an outlet temperature of carbon dioxide gas from the turbine as ~87 °C. The theoretical power produced by the turbine $w = -nM\Delta H$ ($n$ is the molar flowrate of the gas stream; $M$ is the molecular weight of carbon dioxide) is related to the enthalpy change $\Delta H$. The residual enthalpy $H^{res}$ is

$$H^{res}(T,P) = A^{res} + TS^{res} + P^{res}V$$

$$= A^{res} + TS^{res} + RT(Z-1)$$

$$= -RT\ln(1-\rho b) + \frac{a(T)}{2\sqrt{2}b}\ln\left[\frac{1-(\sqrt{2}-1)\rho b}{1+(\sqrt{2}+1)\rho b}\right] + RT\ln(1-\rho b)$$

$$- \frac{T}{2\sqrt{2}b}\frac{\partial a(T)}{\partial T}\ln\left[\frac{1-(\sqrt{2}-1)\rho b}{1+(\sqrt{2}+1)\rho b}\right] + \frac{RT\rho b}{1-\rho b} - \frac{a(T)\rho}{(1+2\rho b-\rho^2 b^2)} \qquad (12)$$

and the enthalpy change of an ideal gas is

$$\Delta H^{ig} = \int_{T_1}^{T_2} C_p^{ig}(T)\, dT \qquad (13)$$

Hence, the theoretical power obtained from the turbine (i.e. 27.8 MW) is comparable with a next generation biomass plant that sources low-grade wood fuel to generate renewable heat and power for a 220-acre park site with CHP steam turbine technology; generating 11-15 MW of electric power and 8-12 MW of heat that is enough to serve 21,000 homes[11].

**Material selection**

Given the complexity of molten salt processes, an empirical approach based upon process experience guides material selection during technology development since selecting an optimum anode material on a theoretical basis is impossible. The application of carbon-based materials in the Hall-Héroult process has detrimental effects on current efficiency and reactor performance[3,12,13]. The same was true for the molten salt Alcoa process for aluminum production in a molten chloride-based salt[6]. The Alcoa reactor had a terminal anode at the top and a terminal cathode at the bottom with 12 to 30 bipolar carbon electrodes stacked vertically at an inter-electrode gap of 1 cm. This reactor had distinct technical advantages over conventional Hall-Héroult reactors: substantially lower temperatures, relatively high current densities, low carbon anode consumption, low fluoride emissions, and smaller footprints. However, the Alcoa process plant was eventually shut down because of a combination of the costs to produce anhydrous aluminum chloride feed, the failure to reach full design capacity, the need to remove and destroy trace amounts of chlorinated biphenal byproducts, the reactor capital costs, and general maintenance costs.

Incorporating inert anodes into the Hall-Héroult process promised a technological revolution in terms of major changes in reactor design and operating practice[6,14]. The bipolar connections used in the Alcoa process allowed electrical energy to be introduced into the reactor by increasing voltage rather than current while generating highly uniform current distributions. It also avoided the scale-up problems associated with conventional monopolar connections while producing an equivalent amount of aluminum. Hence, the production of aluminum could be most efficient in a Hall-Héroult reactor equipped with dimensionally stable inert bipolar electrodes provided that certain technical issues are addressed[15,16]:

- Deployment of electrodes in closely-spaced horizontal arrays to ensure high electrode packing density and so a low capital outlay per unit of aluminum production;

- Anode and cathode materials with dissimilar expansion coefficients within a monolithic structure; minimize ohmic losses;

- Maintaining internal stability of bipolar electrodes during extended reactor operation while protecting the perimeter of the anode/cathode interface from corrosive attack.

The anodic and cathodic side of a bipolar electrode could consist of cooled pipes or flow-channels carrying cooling fluid embedded in the anode while a layer of material (with a higher electrical resistivity than the cathode) is inserted into the bipolar electrode to provide heating[17]. The cathode of the bipolar electrode is heated by means of reducing the active surface area of the cathode so that the electrode has a cathode to anode surface area ratio of 0.5-1. Heating and cooling schedules are imposed to maintain the temperature at the anode and cathode surface slightly lower than the molten salt temperature with careful monitoring of the rate of fluid transport through flow-channels to avoid excessive cooling.

The cathodic protection of Hall-Héroult reactor vessels to withstand the corrosive nature of molten salts is a complex issue and a major challenge in molten salt technology. Matson et al.[18] found that vessels experienced severe corrosive attack in contact with a molten fluoride-based salt. Similarly, Andriiko et al.[19] tested a number of sacrificial anodes to cathodically protect graphite crucibles used in the electrodeposition of germanium from a molten fluoride-based salt. It was not always possible to obtain the required cathodic current density with a graphite crucible anode where degradation of cathode material became oxidized near the anode resulting in product contamination. They also found that the corrosion resistance of copper and nickel was significantly better than iron. Kolosov et al[20] found that imposing an impressed current density provided sufficient corrosion protection of process vessels containing a molten chloride-based salt. However, Ives and Goodman[21] ruled out the use of cathodic protection as it failed to provide corrosion protection of a storage vessel in the vapor regions of a molten carbonate-based salt, and recommended that a lower operating temperature be considered. It is clear that the type of cathodic protection system and molten salt as well as the operating temperature are key variables to consider in the cathodic protection of Hall-Héroult reactor vessels.

### Nuclear application

The incidents and disasters that have marred the nuclear industry strongly indicate a paradigm shift in nuclear fuel technology and power generation. Although oxide fuel (essentially $UO_2$) in the form of cylindrical pellets is most widely used in power reactors, their low fissile atom density, reaction with sodium coolant, and two moderating atoms for each metal atom justifies the need to study alternative fuel types[22]. Also, the thermal conductivity of oxide fuels must be improved to reduce the temperature gradient across the fuel, thereby keeping the center of the fuel well below its melting point ($UO_2$ has a poor thermal conductivity)[1]. It is no surprise that alternative fuel concepts, in particular carbide fuel (Fig. 2), are considered for the next-generation of advanced reactors: high temperature reactors, fast reactors, molten salt reactors[23]. Despite the list of superlatives outlined in Figure 2, key issues for carbide fuel development are in-pile swelling, fission gas retention and the pyrophoric character of fine uranium carbide particles[22].

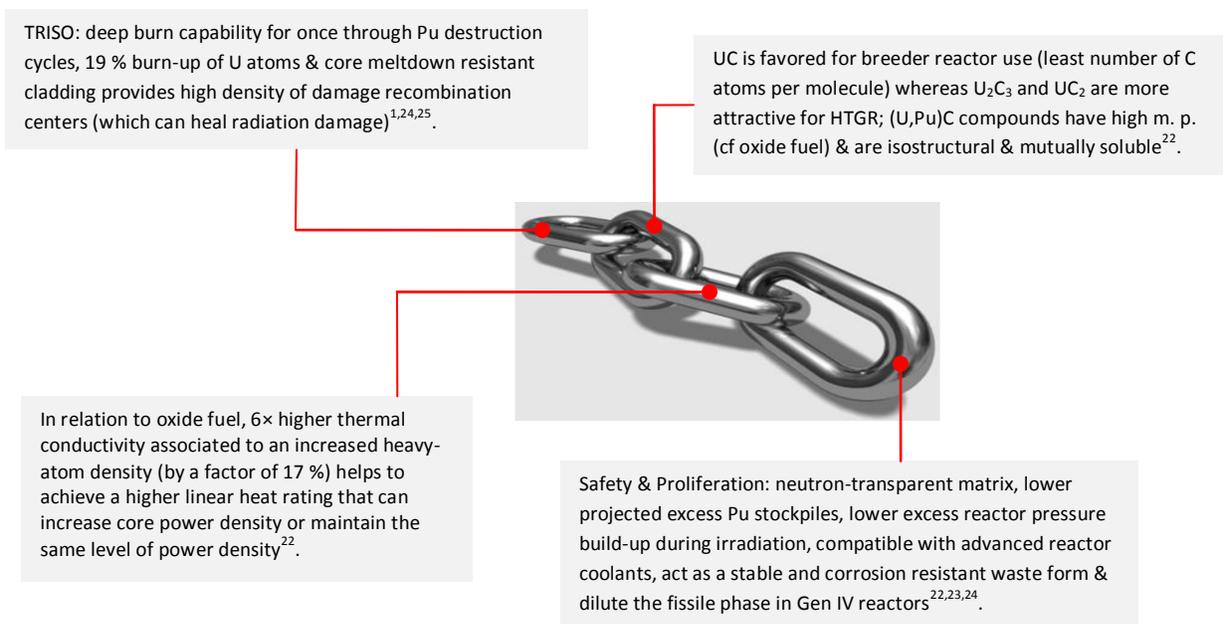

TRISO: deep burn capability for once through Pu destruction cycles, 19 % burn-up of U atoms & core meltdown resistant cladding provides high density of damage recombination centers (which can heal radiation damage)[1,24,25].

UC is favored for breeder reactor use (least number of C atoms per molecule) whereas $U_2C_3$ and $UC_2$ are more attractive for HTGR; (U,Pu)C compounds have high m. p. (cf oxide fuel) & are isostructural & mutually soluble[22].

In relation to oxide fuel, 6× higher thermal conductivity associated to an increased heavy-atom density (by a factor of 17 %) helps to achieve a higher linear heat rating that can increase core power density or maintain the same level of power density[22].

Safety & Proliferation: neutron-transparent matrix, lower projected excess Pu stockpiles, lower excess reactor pressure build-up during irradiation, compatible with advanced reactor coolants, act as a stable and corrosion resistant waste form & dilute the fissile phase in Gen IV reactors[22,23,24].

**Figure 2: Superlatives describing carbide fuel in advanced Generation IV reactor concepts.**

To achieve a global electricity system that is sustainable over the first half of this century, proliferation resistant weapon technologies and safe nuclear waste treatment is vital. The encapsulation and direct disposal of plutonium from spent nuclear fuel has been conceived as the less costly and less risky method of lowering excess global plutonium stockpiles[26,27,28,29]. However, simply burying plutonium in deep geologic repositories would disregard the energy it could generate and make future handling of the stockpile much more difficult. In constrast, molten salt reprocessing has an emerged as a viable long term option: able to recycle and co-recover actinides in a single process with a relatively low amount of liquid waste, reduce the long term heat production and HLW volume, radiation stability of molten salts, and proliferation resistance[30,31]. An understanding of molten salt reprocessing of spent nuclear fuel in a sustainable closed loop cycle is vital in the development of advanced reactors[32]. Hence, the inclusion of carbon dioxide cycles in the molten salt reprocessing of carbide fuel could offer another route for electricity generation.

## Future challenges

**Marketing.** The marketing strategy of micro-market segmentation and network-partnership relationships play crucial roles in persuasion and influencing public perception about the application of molten salt technology in electricity generation. For instance, molten-salt reactors have the potential to solve almost all the problems of nuclear energy in a far more elegant way than existing light-water reactors[23]. However, relying on public input to guide the development of a new technology may be counter-productive as they have difficulty in articulating needs beyond the realm of their own experiences[33,34]. Although public preference data of e-marketing reduces costs in relation to the tedious generation of psychographic data (required by network-partnership relationships) caution is needed to avoid excessive e-marketing. In the highly turbulent e-market environment, arising from rapid technological advances, market uncertainties and intense risk, make marketing efforts even more important to the success of a new technology. This suggests that a synergy of e-marketing and network-partnership relationships (Fig. 3) is needed for the introduction and survival of molten salt technology in future power generation markets.

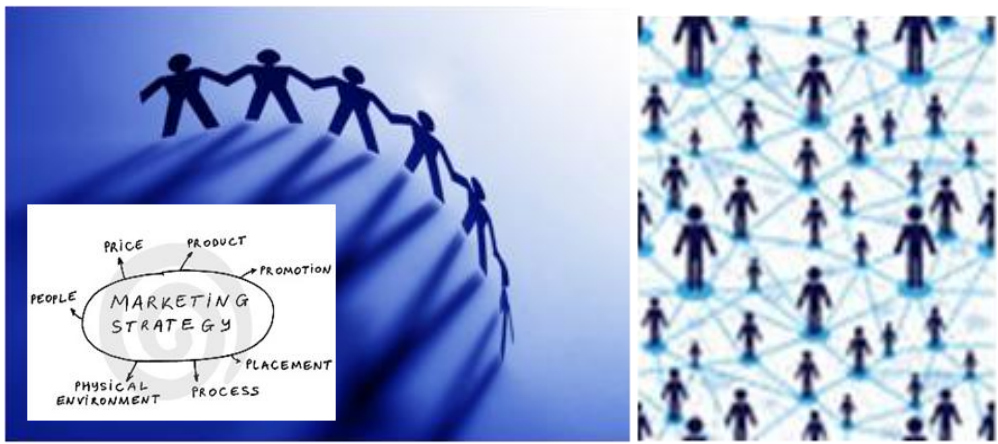

**Figure 3: Marketing strategy applied to e-marketing and network-partnership relationships**

**Human resources.** The role of regulatory safety inspectors could address negative perceptions of job communication and work environment as these are predictive of high-accident, unsafe behaviors and negative safety cultures. Their role will be vitally important in the development of the next generation of advanced reactors since the nuclear skills base is aging and lacks regulatory safety inspectors[1]. Therefore, the role of human resource management within the industry should focus on employee interaction and empowerment through the delegation of authority by management and their participation in the decision-making process[35]. In particular, this would integrate human resource and management practices (e.g. employee of the month, reward pay for good safety behavior, positive behavior awards, promotion systems and so on) in order to foster motivation, safe behavior and personal accountability[36,37]. Such systems should be regularly monitored and reviewed through employee feedback to consider the amount of pay awarded for different levels of performance (amount of pay that any employee or team receives over another employee or team), and should identify clear criteria between cash and non-cash awards.

**Acknowledgements** I thank J. Osarinmwian for discussions and inspiration. This research was funded by EPSRC (UK).